\documentclass[twocolumn]{aastex62}

\usepackage{amsmath}
\usepackage[Symbol]{upgreek}
\usepackage{bm}
\usepackage[bookmarks=false,         
     pdfnewwindow=true,      
     colorlinks=true,    
     linkcolor=xlinkcolor,     
     citecolor=xlinkcolor,     
     filecolor=xlinkcolor,  
     urlcolor=xlinkcolor,      
     final=true
]{hyperref}
\usepackage[capitalise]{cleveref}
\crefname{subsection}{subsection}{subsections}

\received{XXX}
\revised{XXX}
\accepted{XXX}

\shorttitle{Measurements of tropospheric ice clouds with Polarbear}
\shortauthors{Takakura et al.}

\begin{document}

\title{Measurements of tropospheric ice clouds with a ground-based CMB polarization experiment, \textsc{Polarbear}}

\correspondingauthor{Satoru Takakura}
\email{satoru.takakura@ipmu.jp}

\author[0000-0001-9461-7519]{S. Takakura}
\affiliation{Kavli IPMU (WPI), UTIAS, The University of Tokyo, Kashiwa, Chiba 277-8583, Japan}
\author[0000-0002-1571-663X]{M.~A.~O. Aguilar-Fa\'undez}
\affiliation{Departamento de F\'isica, FCFM, Universidad de Chile, Blanco Encalada 2008, Santiago, Chile}
\affiliation{Department of Physics and Astronomy, Johns Hopkins University, Baltimore, MD 21218, USA}
\author{Y. Akiba}
\affiliation{Department of Particle and Nuclear Physics, SOKENDAI, Hayama, Kanagawa 240-0193, Japan}
\author[0000-0002-3407-5305]{K. Arnold}
\affiliation{Department of Physics, University of California, San Diego, CA 92093-0424, USA}
\author{C. Baccigalupi}
\affiliation{International School for Advanced Studies (SISSA), Via Bonomea 265, 34136, Trieste, Italy}
\affiliation{The National Institute for Nuclear Physics, INFN, Sezione di Trieste Via Valerio 2, I-34127, Trieste, Italy}
\author[0000-0002-1623-5651]{D. Barron}
\affiliation{Department of Physics and Astronomy, University of New Mexico, Albuquerque, NM 87131,  USA}
\author{D. Beck}
\affiliation{AstroParticule et Cosmologie (APC), Univ Paris Diderot, CNRS/IN2P3, CEA/Irfu, Obs de Paris, Sorbonne Paris Cit\'e, France}
\author[0000-0003-4847-3483]{F. Bianchini}
\affiliation{School of Physics, University of Melbourne, Parkville, VIC 3010, Australia}
\author{D. Boettger}
\affiliation{Instituto de Astrof\'isica and Centro de Astro-Ingenier\'ia, Facultad de F\'isica, Pontificia Universidad Cat\'olica de Chile, Vicu\~na Mackenna 4860, 7820436 Macul, Santiago, Chile}
\author{J. Borrill}
\affiliation{Computational Cosmology Center, Lawrence Berkeley National Laboratory, Berkeley, CA 94720, USA}
\author{K. Cheung}
\affiliation{Department of Physics, University of California, Berkeley, CA 94720, USA}
\author[0000-0002-3266-857X]{Y. Chinone}
\affiliation{Department of Physics, University of California, Berkeley, CA 94720, USA}
\affiliation{Kavli IPMU (WPI), UTIAS, The University of Tokyo, Kashiwa, Chiba 277-8583, Japan}
\author{T. Elleflot}
\affiliation{Department of Physics, University of California, San Diego, CA 92093-0424, USA}
\author{J. Errard}
\affiliation{AstroParticule et Cosmologie (APC), Univ Paris Diderot, CNRS/IN2P3, CEA/Irfu, Obs de Paris, Sorbonne Paris Cit\'e, France}
\author[0000-0002-3255-4695]{G. Fabbian}
\affiliation{Institut d'Astrophysique Spatiale, CNRS (UMR 8617), Univ.~Paris-Sud, Universit\'e Paris-Saclay, b\^at.\ 121, 91405 Orsay, France}
\author{C. Feng}
\affiliation{Department of Physics, University of Illinois at Urbana-Champaign, Urbana, IL 61801, USA}
\author{N. Goeckner-Wald}
\affiliation{Department of Physics, University of California, Berkeley, CA 94720, USA}
\author{T. Hamada}
\affiliation{Astronomical Institute, Graduate School of Science, Tohoku University, Sendai, Miyagi 980-8578, Japan}
\author{M. Hasegawa}
\affiliation{Institute of Particle and Nuclear Studies, High Energy Accelerator Research Organization (KEK), Tsukuba, Ibaraki 305-0801, Japan}
\author{M. Hazumi}
\affiliation{Institute of Particle and Nuclear Studies, High Energy Accelerator Research Organization (KEK), Tsukuba, Ibaraki 305-0801, Japan}
\affiliation{Department of Particle and Nuclear Physics, SOKENDAI, Hayama, Kanagawa 240-0193, Japan}
\affiliation{Kavli IPMU (WPI), UTIAS, The University of Tokyo, Kashiwa, Chiba 277-8583, Japan}
\affiliation{Institute of Space and Astronautical Science, Japan Aerospace Exploration Agency (JAXA), Sagamihara, Kanagawa 252-0222, Japan}
\author{L. Howe}
\affiliation{Department of Physics, University of California, San Diego, CA 92093-0424, USA}
\author{D. Kaneko}
\affiliation{Kavli IPMU (WPI), UTIAS, The University of Tokyo, Kashiwa, Chiba 277-8583, Japan}
\author{N. Katayama}
\affiliation{Kavli IPMU (WPI), UTIAS, The University of Tokyo, Kashiwa, Chiba 277-8583, Japan}
\author[0000-0003-3118-5514]{B. Keating}
\affiliation{Department of Physics, University of California, San Diego, CA 92093-0424, USA}
\author[0000-0001-5748-5182]{R. Keskitalo}
\affiliation{Computational Cosmology Center, Lawrence Berkeley National Laboratory, Berkeley, CA 94720, USA}
\author{T. Kisner}
\affiliation{Computational Cosmology Center, Lawrence Berkeley National Laboratory, Berkeley, CA 94720, USA}
\affiliation{Space Sciences Laboratory, University of California, Berkeley, CA 94720, USA}
\author{N. Krachmalnicoff}
\affiliation{International School for Advanced Studies (SISSA), Via Bonomea 265, 34136, Trieste, Italy}
\author{A. Kusaka}
\affiliation{Physics Division, Lawrence Berkeley National Laboratory, Berkeley, CA 94720, USA}
\affiliation{Department of Physics, The University of Tokyo, Bunkyo-ku, Tokyo 113-0033, Japan}
\author[0000-0003-3106-3218]{A.~T. Lee}
\affiliation{Department of Physics, University of California, Berkeley, CA 94720, USA}
\affiliation{Physics Division, Lawrence Berkeley National Laboratory, Berkeley, CA 94720, USA}
\author{L.~N. Lowry}
\affiliation{Department of Physics, University of California, San Diego, CA 92093-0424, USA}
\author{F.~T. Matsuda}
\affiliation{Kavli IPMU (WPI), UTIAS, The University of Tokyo, Kashiwa, Chiba 277-8583, Japan}
\author{A.~J. May}
\affiliation{Jodrell Bank Centre for Astrophysics, University of Manchester, Manchester, M13~9PL, UK}
\author[0000-0003-2176-8089]{Y. Minami}
\affiliation{Institute of Particle and Nuclear Studies, High Energy Accelerator Research Organization (KEK), Tsukuba, Ibaraki 305-0801, Japan}
\author{M. Navaroli}
\affiliation{Department of Physics, University of California, San Diego, CA 92093-0424, USA}
\author[0000-0003-0738-3369]{H. Nishino}
\affiliation{Institute of Particle and Nuclear Studies, High Energy Accelerator Research Organization (KEK), Tsukuba, Ibaraki 305-0801, Japan}
\author{L. Piccirillo}
\affiliation{Jodrell Bank Centre for Astrophysics, University of Manchester, Manchester, M13~9PL, UK}
\author{D. Poletti}
\affiliation{International School for Advanced Studies (SISSA), Via Bonomea 265, 34136, Trieste, Italy}
\author[0000-0002-0689-4290]{G. Puglisi}
\affiliation{Department of Physics and KIPAC, Stanford University, Stanford, CA 94305, USA}
\author[0000-0003-2226-9169]{C.~L. Reichardt}
\affiliation{School of Physics, University of Melbourne, Parkville, VIC 3010, Australia}
\author{Y. Segawa}
\affiliation{Department of Particle and Nuclear Physics, SOKENDAI, Hayama, Kanagawa 240-0193, Japan}
\author[0000-0001-7480-4341]{M. Silva-Feaver}
\affiliation{Department of Physics, University of California, San Diego, CA 92093-0424, USA}
\author[0000-0001-6830-1537]{P. Siritanasak}
\affiliation{Department of Physics, University of California, San Diego, CA 92093-0424, USA}
\author{A. Suzuki}
\affiliation{Physics Division, Lawrence Berkeley National Laboratory, Berkeley, CA 94720, USA}
\author{O. Tajima}
\affiliation{Department of Physics, Kyoto University, Kyoto, Kyoto 606-8502, Japan}
\author{S. Takatori}
\affiliation{Institute of Particle and Nuclear Studies, High Energy Accelerator Research Organization (KEK), Tsukuba, Ibaraki 305-0801, Japan}
\author{D. Tanabe}
\affiliation{Department of Particle and Nuclear Physics, SOKENDAI, Hayama, Kanagawa 240-0193, Japan}
\author{G.~P. Teply}
\affiliation{Department of Physics, University of California, San Diego, CA 92093-0424, USA}
\author{C. Tsai}
\affiliation{Department of Physics, University of California, San Diego, CA 92093-0424, USA}

\begin{abstract}
The polarization of the atmosphere has been a long-standing concern for ground-based experiments targeting cosmic microwave background (CMB) polarization.
Ice crystals in upper tropospheric clouds scatter thermal radiation from the ground and produce a horizontally-polarized signal.
We report the detailed analysis of the cloud signal using a ground-based CMB experiment, \textsc{Polarbear}, located at the Atacama desert in Chile and observing at 150\,GHz.
We observe horizontally-polarized temporal increases of low-frequency fluctuations (``polarized bursts,'' hereafter) of $\lesssim0.1\,\mathrm{K}$ when clouds appear in a webcam monitoring the telescope and the sky.
The hypothesis of no correlation between polarized bursts and clouds is rejected with $>24\,\sigma$ statistical significance using three years of data.
We consider many other possibilities including instrumental and environmental effects, and find no other reasons other than clouds that can explain the data better.
We also discuss the impact of the cloud polarization on future ground-based CMB polarization experiments.
\end{abstract}

\keywords{atmospheric effects --- scattering --- cosmology: observations --- cosmic background radiation --- polarization}

\section{Introduction}\label{sec:introduction}
The atmosphere is an unavoidable foreground in any measurement with a ground-based telescope.
Absorption, emission, and scattering by atmospheric molecules define the exploitable wavelength windows for astronomical observations.
In addition, turbulence in the troposphere due to convective heat transfer causes variable weather conditions and reduces the observing efficiency.

In particular, cosmic microwave background (CMB) experiments observe the sky for thousands of hours to measure very faint anisotropies from the early universe, such as degree-scale parity-odd ($B$-mode) polarization anisotropies generated by primordial gravitational waves~\citep{SeljakZaldarriaga1997PhRvL}.
Atmospheric fluctuations introduce gradually varying (low-frequency) noise and degrade the CMB anisotropy measurements at large-angular scales~\citep{LayHalverson2000ApJ}.
Therefore, the polarization of the atmosphere is a very significant concern for current and future ground-based CMB experiments.

The atmospheric transmission windows for CMB observation are typically ${<}50$, 70--110, 120--180, and 190--320\,GHz bands.
The atmospheric emission in this frequency range is dominated by oxygen and water vapor~\citep[e.g.][]{Westwater2004IEEE}.
Fortunately, the emission is almost completely unpolarized~\citep{ABS2014Demod,Errard2015ApJ}, or slightly circularly polarized because of Zeeman splitting due to the Earth's magnetic field~\citep{Rosenkranz1988RaSc,Keating1998ApJ,HananyRosenkranz2003NewAR,Spinelli2011MNRAS}.
Although density and temperature fluctuations in the turbulent atmosphere cause significant low-frequency noise for CMB intensity (or temperature) measurements, they do not affect linear polarization measurements if the instrumental polarization leakage is negligible.

However, clouds in the atmosphere could produce linearly polarized microwave radiation.
Clouds consist of small ice crystals, water droplets, or both depending on atmospheric conditions, and these small particles scatter the thermal radiation mainly coming from the ground.
The scattered light appears as a horizontally-polarized signal in the line of sight~\citep{Troitsky2000R&QE,Pietranera2007MNRAS}.
Furthermore, the horizontal alignment of ice crystals having a column or plate shape~\citep{Ono1969JAtS,Chepfer1999JQSRT} increases the polarization signal~\citep{Czekala1998GeoRL}.
This linearly-polarized signal from anisotropic clouds is a source of low-frequency noise for linear polarization data.
It cannot be mitigated even with ideal instruments or by other techniques such as polarization modulation~\citep{Brown2009}.

The impact of the polarized signal from clouds for CMB polarization measurements is fully discussed in \cite{Pietranera2007MNRAS} and partially mentioned in \cite{Kuo2017ApJ}.
Measurements of the signal have been reported in the atmospheric science community using microwave radiometers~\citep{Troitsky2000R&QE,Troitsky2003JAtS,Troitsky2005R&QE,Kneifel2010JGRD,Xie2012JGRD,Defer2014JGRD,Xie2015JGRD,Pettersen2016ACP,GongWu2017ACP}.
In the CMB community, the BICEP2 Collaboration~(\citeyear{BICEP2experiment2014ApJ}) mentions the possibility of low-frequency noise ($1/f$ noise) from clouds and ABS~\citep{ABS2018} reports the existence of noise flare-ups in the polarization signal.

In this paper, we report measurements of the polarization of clouds at \textsc{Polarbear}, a ground-based experiment observing CMB polarization at 150\,GHz from the Atacama Desert in Chile.
To our knowledge, this is the first detailed report of this kind of effect using a CMB instrument.
One of the unique features of \textsc{Polarbear} is polarization modulation using a continuously-rotating half-wave plate (CRHWP)~\citep{Takakura2017JCAP,Hill2016SPIE}.
This technique mitigates the spurious polarization due to the leakage of unpolarized signals and instrumental temperature variations, and we can clearly measure the polarization from the sky. 

In \autoref{sec:model}, we briefly explain the basics of the scattering of microwave radiation by ice crystals within the clouds. 
In \autoref{sec:measurements}, we show an example of polarization measurements during a cloudy day and then look for similar observations in 2.5 years of data.
Following the results, we discuss the impact of the clouds on the CMB experiments in \autoref{sec:discussion} and summarize this study in \autoref{sec:summary}.

\section{Basics of ice clouds} \label{sec:model}
The \textsc{Polarbear} experiment is located at the James Ax Observatory, at an altitude of $5{,}200\,\mathrm{m}$ on Cerro Toco.
This site, in the Atacama Desert in Northern Chile, is in one of the driest regions on the Earth. However, clouds still do occasionally exist there \citep[e.g.\ Fig.~7 of ][]{Kuo2017ApJ}.
Clouds form when a moist air parcel is adiabatically expanded due to a rapid change in elevation and its water vapor content supersaturates.
Clouds take various forms, which are typically classified into ten types depending on the atmospheric condition~\citep[e.g.][]{LiouYang2016CUP}.
High clouds (cirrus, cirrocumulus, and cirrostratus) form at altitudes around 5,000--$13{,}000\,\mathrm{m}$ and are therefore the most relevant to observations at \textsc{Polarbear}'s altitude~\citep{Erasmus2001CTIO}.

The high clouds consist mainly of ice crystals,%
\footnote{We focus on ice clouds in this study, but the same model with different parameters can be applied to clouds that consist of water droplets. Since the water droplets are spherical and more absorptive than ice, the polarization fraction would be small.}
which have various properties depending on the atmospheric condition, i.e.\ temperature and water vapor content, as well as the evolution of the clouds.
The mean effective size of an ice crystal is typically $D_e \simeq20$--$100\,\mu\mathrm{m}$. 
The ice water content (IWC), which is the density of ice in the clouds, is about $10^{-3}$--$10^{-1}\,\mathrm{g}\cdot\mathrm{m}^{-3}$~\citep{Rolland2000JGR}.
This results in a number density $n$ of about $10^{4}$--$10^{5}\,\mathrm{m}^{-3}$.
The ice water path (IWP), which is the total mass of ice crystals per unit area, is about $1$--$10\,\mathrm{g}\cdot\mathrm{m}^{-2}$~\citep{Kuo2017ApJ}.
Thus the geometric thickness of the cirrus clouds $\Delta h$ is about $10^{3}\,\mathrm{m}$.

Ice crystals in clouds are not spherical.
Small, primary crystals take the form of a hexagonal column and evolve to longer columns, larger hexagonal plates, or their aggregates.
The majority of ice crystals take the hexagonal column shape, which is measured by \cite{Ono1969JAtS} and \cite{Chepfer1999JQSRT}.
The aspherical shapes could cause alignment of the ice crystals due to the drag of the atmosphere~\citep{Ono1969JAtS}.
One can see the signal as characteristic halos such as sundogs, circumzenithal/circumhorizontal arcs, and upper/lower tangent arcs~\citep{HaloSim}.

\subsection{Rayleigh Scattering}\label{subsec:Rayleigh}
Since the size of ice crystals is sufficiently smaller than the wavelength, the scattering of microwave radiation by ice crystals is described by Rayleigh scattering.
The electric field $\bm{E}_\mathrm{sc}$ scattered by an ice crystal located at the origin is expressed as \citep[e.g.][]{LandauLifshitz1960}
\begin{equation}\label{eq:dipoleemission}
\bm{E}_\mathrm{sc}(r\hat{\bm{n}}) = - \frac{\omega^2}{c^2 r}\hat{\bm{n}}\times(\hat{\bm{n}}\times{\bm{P}})\;,
\end{equation}
where $r$ is the distance from the origin, $\hat{\bm{n}}$ is a unit vector toward the propagation direction, $\omega$ is the angular frequency of the wave, and $c$ is the speed of light.
In general, the electric dipole moment $\bm{P}$ is expressed as
\begin{equation}\label{eq:dipolemoment}
\bm{P} = V\bm{\upalpha}\bm{E}_\mathrm{in}\;,
\end{equation}
where $V$ is the volume of the scatterer, $\bm{E}_\mathrm{in}$ is the incident electric field, and $\bm{\upalpha}$ is the polarizability matrix, calculated as
\begin{equation}\label{eq:polarizability}
\bm{\upalpha} = \frac{\epsilon-1}{4\pi}\Big[\bm{\mathrm{I}}+(\epsilon-1)\bm{\Updelta}\Big]^{-1}\;,
\end{equation}
where $\bm{\mathrm{I}}$ is the identity matrix and $\bm{\Updelta}$ is the depolarization factor, which is a positive definite symmetric matrix satisfying $\mathrm{Tr}\,(\bm{\Updelta})=1$ and depends on the shape and orientation of the scatterer.
In the case of spheroids, 
the $\bm{\Updelta}$ is parameterized as a diagonal matrix $\mathrm{diag}\{(1-\Delta_z)/2,\,(1-\Delta_z)/2,\,\Delta_z\}$, where $\Delta_z<1/3$, $\Delta_z=1/3$, and $\Delta_z>1/3$ correspond to the prolate (column), spherical, and oblate (plate) shapes, respectively.
The relative permittivity $\epsilon$ of ice is about $3.15$ for microwave radiation~\citep{WarrenBrandt2008JGRD}.

In the simple case of spherical particles, we can obtain the total cross-section of the Rayleigh scattering $\sigma_\mathrm{R}$ as
\begin{equation}\label{eq:opticaldepth}
\sigma_\mathrm{R} = \frac{8\pi}{3}\frac{V^2\alpha^2\omega^4}{c^4}\propto D_{e}^6\omega^{4}\;,
\end{equation}
where 
$\alpha=(3/4\pi)(\epsilon-1)/(\epsilon+2)$.
We can see the well-known dependence on the size of the scatterer $D_e$ and the frequency of the light $\omega$.
This strong dependence on the particle size and variation of the size distribution in clouds cause huge uncertainty in the prediction of $\sigma_\mathrm{R}$ by orders of magnitude.
At the observing frequency of the \textsc{Polarbear}, $\omega/(2\pi)=150\,\mathrm{GHz}$, the cross-section $\sigma_\mathrm{R}$ becomes $\sim10^{-16}\,\mathrm{m}^2$ for $D_e=20\,\mu\mathrm{m}$ and $\sim10^{-12}\,\mathrm{m}^2$ for $D_e=100\,\mu\mathrm{m}$.
By assuming that all the particles in a cloud have the same size and using the typical number density of ice crystals, $n\sim10^{4}$--$10^{5}\,\mathrm{m}^{-3}$, and the typical thickness of cirrus clouds, $\Delta h\sim10^3\,\mathrm{m}$, we estimate the optical depth of the clouds as $\tau\sim10^{-9}$--$10^{-4}$.
Note that the estimate increases for $220$ or $280\,\mathrm{GHz}$ due to the frequency dependence.

The calculation above has been substantially simplified by ignoring the size and shape distributions but does indicate that larger ice crystals in clouds are the main contributor to the scattering of microwave radiation and that the optical depth would be sub-percent level at most.

However, scattering by clouds changes the direction of the thermal radiation from the ground and injects it into the line of sight.
Since almost half of the solid angle as seen from the cloud is covered by the ground at ambient temperature, scattering of the ground emission at sub-percent levels could cause additional signal at the ${\sim}1\,\mathrm{K}$ level.
The clouds are randomly distributed in the sky and are gradually varying and moving due to atmospheric turbulence and wind, which leads to low-frequency variations of the signal from the clouds.
Therefore, clouds can become an important source of low-frequency noise (see \autoref{sec:discussion} for further discussions).
The cloud signal is significantly larger than the current detector noise level for CMB observations, which is lower than $1\,\mathrm{mK}$ over a few seconds of beam-crossing time, and thus can be detected instantaneously.

\subsection{Polarization of Ice Clouds}\label{subsec:polarizationmodel}
There are two types of effects that polarize the light scattered by the ice crystals.
The first is due to the curvature of the ground.
The second is due to the horizontal alignment of ice crystals with plate or column shape.
We explain the two effects in the following.
Our estimate and observation suggest the latter is dominant.

\begin{figure}[t]\centering
\includegraphics[width=0.4\textwidth]{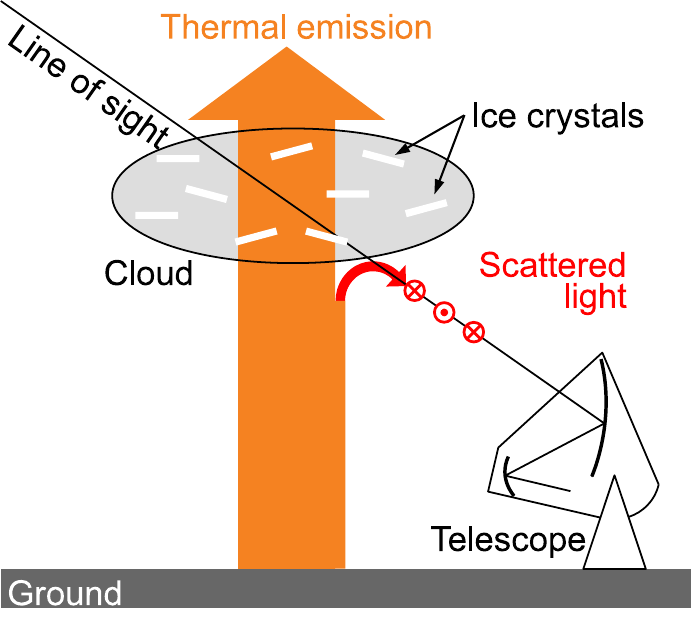}
\caption{\label{fig:cartoon}Illustration of the microwave signal from clouds. Ice crystals scatter thermal emission from the ground and generate horizontal polarization.}
\end{figure}

\begin{figure}[t]\centering
\includegraphics[width=0.45\textwidth]{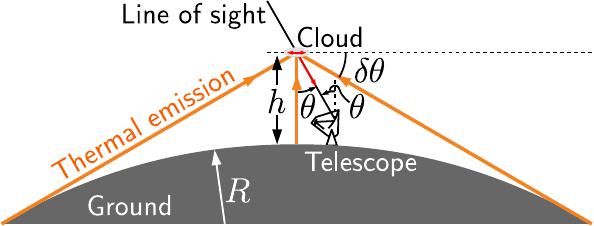}
\caption{\label{fig:earthcurvature}Sketch showing microwave radiation emitted from the spherical ground, reaching a cloud above the observing site.}
\end{figure}

The three-dimensional positional relations among the telescope, the clouds, and the ground produce polarization (\cref{fig:cartoon}).%
As shown in \cref{eq:dipoleemission,eq:dipolemoment}, the polarization of the scattered light is determined by its scattering angle and the polarization of the incident light.
By taking spherical coordinates $(\theta, \phi)$ centered at the clouds and aligning the $z$-axis to zenith as shown in \cref{fig:earthcurvature},%
the ground radiation can be expanded in spherical harmonics $Y_{l}^{m}(\theta, \phi)$ with $m=0$ because of the axial symmetry.
If we assume that the ground is a blackbody with a uniform temperature $T_\mathrm{g}$, the expansion coefficients $a_{l,0}$ are obtained as 
\begin{equation}\label{eq:sphericalexpansion}
a_{l,0} = 2\pi T_\mathrm{g}\int_{0}^{\frac{\pi}{2}-\delta\theta}Y_{l}^{0}(\theta,0)\sin\theta\,d\theta\;.
\end{equation}
Here, $\delta\theta$ is the look-down angle of the horizon from the clouds, which is approximately $\sqrt{2h/R}$, with the altitude of the clouds $h$ and the radius of the Earth $R$.
In particular, the monopole ($l=0$) and quadrupole ($l=2$) components are calculated as
\begin{equation}
\frac{a_{0,0}}{\sqrt{4\pi}T_\mathrm{g}} = \frac{1-\sin\delta\theta}{2}\quad\mathrm{and}\quad
\frac{a_{2,0}}{\sqrt{4\pi}T_\mathrm{g}} = \frac{\sqrt{5}}{4}\sin\delta\theta\cos^2\!\delta\theta\;,
\end{equation}
respectively.
Again, if we assume spherical particles to deal with the scattering simply by the optical depth $\tau$, the Stokes parameters of the scattered light are calculated as~\citep{HuWhite1997PhRvD}
\begin{equation}
\left(\begin{matrix}
I\\Q
\end{matrix}\right)
\approx
\frac{\tau T_\mathrm{g}}{\sqrt{4\pi}} \left(\begin{matrix}
1\\0
\end{matrix}\right) a_{0,0} + 
\frac{\tau T_\mathrm{g}}{10} \left(\begin{matrix}
Y_{2}^{0}(\theta,\phi)\\-\sqrt{6}\!\!\!\!\!{\phantom{Y}}_2^{\phantom{0}}\!Y_{2}^{0}(\theta,\phi)
\end{matrix}\right) a_{2,0}\;,
\end{equation}
where $Y_2^{0}(\theta,\phi)\equiv\sqrt{5/(16\pi)}(3\cos^2\!\theta-1)$, $\!\!\!\!\!{\phantom{Y}}_2^{\phantom{0}}\!Y_{2}^{0}(\theta,\phi)\equiv\sqrt{15/(32\pi)}\sin^2\!\theta$ and the other Stokes parameters, $U$ and $V$, are zero.
Here, the polarizations are defined on the usual base vectors ($\bm{e}_\theta$, $\bm{e}_\phi$), and the negative Stokes $Q$ represents horizontal linear polarization.
The polarization fraction $p$ is the ratio of $Q$ to $I$, thus
\begin{equation}\label{eq:polfrac}
|p|\approx\frac{3}{4\sqrt{5}}\frac{a_{2,0}}{a_{0,0}}\sin^2\!\theta\approx\frac{3}{4}\sqrt{\frac{h}{2R}}\sin^2\!\theta\;.
\end{equation}
Putting $h=10\,\mathrm{km}$, $R=6{,}400\,\mathrm{km}$, and $\theta=45^\circ$ into \cref{eq:polfrac} results in $|p|\sim1\%$.

\begin{figure}[t]\centering
\includegraphics[width=0.5\textwidth]{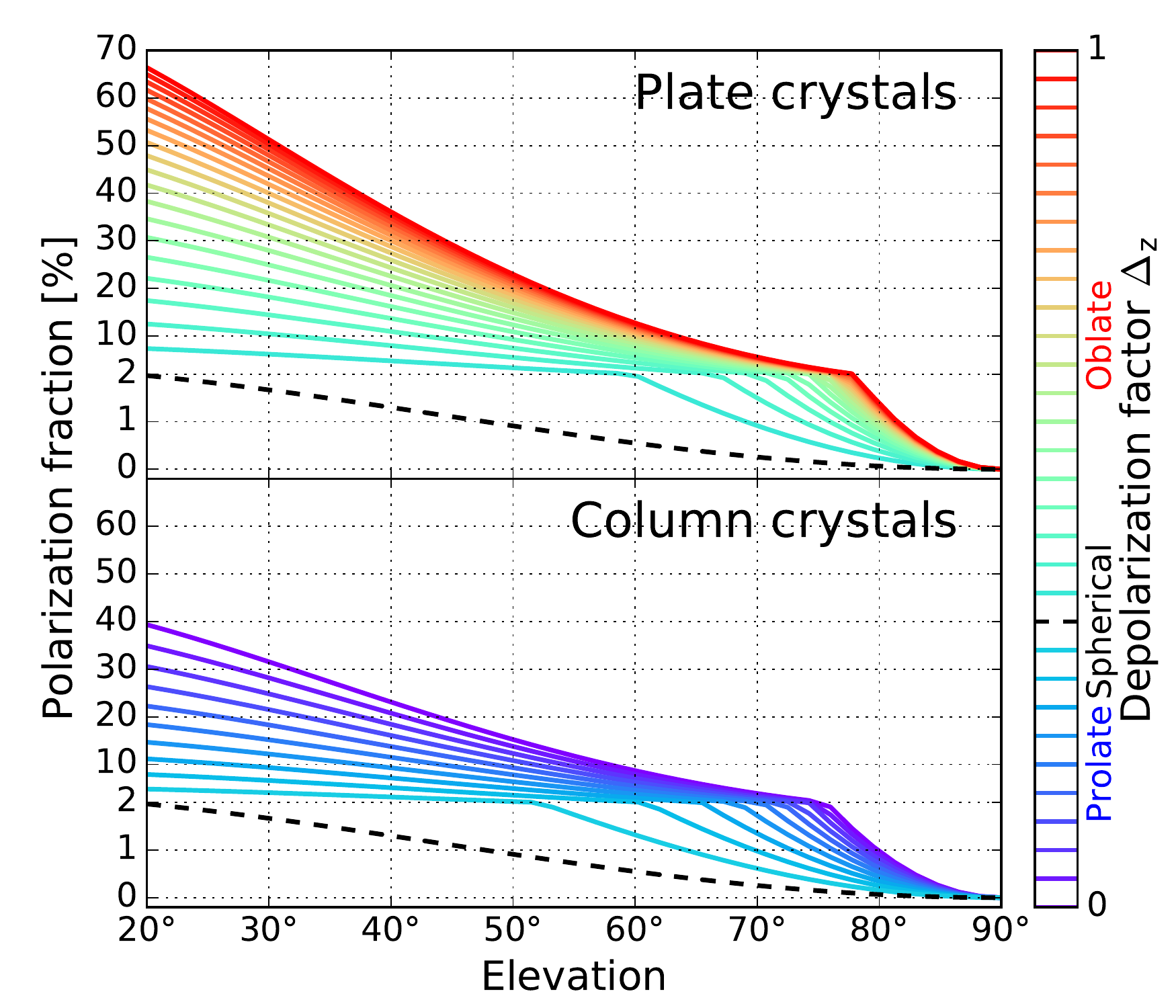}
\caption{\label{fig:polfrac_vs_shape}Calculated polarization fraction of the light scattered by horizontally aligned ice crystals with a spheroidal shape as a function of the elevation.
The dashed black lines show the spherical case.
The line color of the other lines represents the shape information parameterizing the $z$-component of the depolarization factor $\Delta_z$.
The column case shows the averaged contribution among randomly oriented particles in azimuth.}
\end{figure}

Horizontally aligned ice crystals with column and plate shapes scatter horizontal electric fields more efficiently and can, therefore, produce a larger polarized signal.
We approximate the crystal shapes as spheroids and directly calculate \cref{eq:dipoleemission,eq:dipolemoment,eq:polarizability}.
In the case of column shape, we set the long axis horizontal but assume that its azimuth is random.
The polarization fraction calculated from the model is shown in \cref{fig:polfrac_vs_shape}.%
Here, all the crystals are assumed to have the same shape and to be horizontally aligned, i.e.\ the long axis of the spheroid is in the horizontal plane.
Thus, these estimates give an upper limit, whereas real polarization fractions are expected to be smaller.
However, the amplitude of the polarization fraction is considerably larger than the spherical case shown as the black line.
This is because the horizontally aligned crystals have a tendency to scatter horizontally polarized light even without a quadrupolar anisotropy of the incident radiation field.

\section{Measurements}\label{sec:measurements}
We analyze the data taken by \textsc{Polarbear}~\citep{POLARBEAR2017ApJ} and search for signals that appear to be from tropospheric ice clouds.
We use transition-edge sensor (TES) bolometers in the \textsc{Polarbear} receiver~\citep{POLARBEAR2012SPIEKam} and a webcam monitoring 
the exterior of \textsc{Polarbear} and its surroundings, including the sky.
Each TES bolometer is coupled to a dipole-slot antenna and measures a single polarization of the incident light.
In addition, \textsc{Polarbear} has a CRHWP at the prime focus~\citep{Takakura2017JCAP,Hill2016SPIE}, which is continuously rotated at $2\,\mathrm{Hz}$ to modulate the polarization signal from the sky.
Thus, the modulated timestream of the detector $d_m(t)$ is expressed as
\begin{equation}
d_m(t) = I(t) + \mathrm{Re}\{[A_0 + Q(t)+iU(t)]e^{-i\omega_\mathrm{m}t}\}\;,
\end{equation}
where $I(t)$, $Q(t)$ and $U(t)$ are variations of the Stokes parameters of the sky,
$A_0$ is the steady polarization from the instruments, and $\omega_\mathrm{m}$ is the modulation frequency.
Throughout the following analysis, the Stokes $Q(t)$ and $U(t)$ are defined on the instrumental coordinates ($\bm{e}_\mathrm{ZE},\bm{e}_\mathrm{AZ}$), where ZE and AZ represent the zenith and azimuth angle, respectively.
We demodulate the timestream and extract the polarization signal as a demodulated timestream $d_d(t)$:
\begin{equation}
d_d(t) = A_0 + Q(t)+iU(t)\;.
\end{equation}
We can also obtain the intensity signal by applying a low-pass filter.
See \cite{Takakura2017JCAP} for more details.
For the analysis in \autoref{subsec:example}, no filters are applied anymore because the signal is very significant.
For the analysis in \autoref{subsec:bolostat}, we apply the polynomial filter, the scan-synchronous signal filter, and the intensity-to-polarization leakage filter~\citep{Takakura2017JCAP} to mitigate spurious contributions such as the responsivity variation of the detectors, polarized ground signals, and instrumental polarization.
For the dataset used in this paper, the absolute polarization angle calibration is still preliminary and the calibration uncertainty is about a few degrees.
While further work is in progress toward the final calibration for CMB science analysis, this preliminary calibration suffices for the purpose of the study presented in this paper.

The webcam is placed in the control container located $17.2\,\mathrm{m}$ north of the telescope.
Since it is mainly used to monitor the telescope, its field of view (FOV) covers about $130^\circ$--$180^\circ$ in azimuth and $-10^\circ$--$20^\circ$ in elevation, and it takes a picture every $5$ minutes.

\subsection{Example}\label{subsec:example}
\begin{figure}\centering
\includegraphics[width=0.5\textwidth]{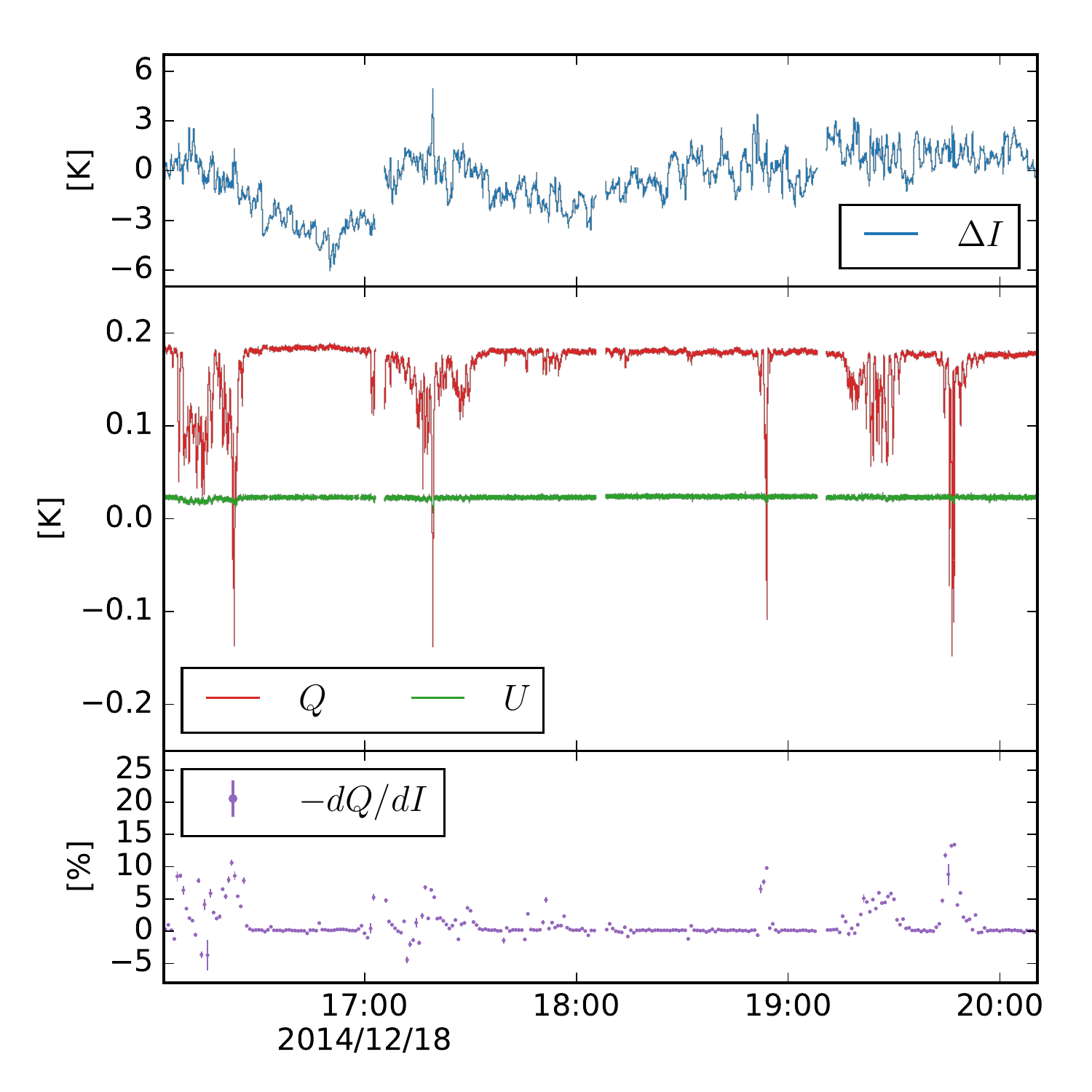}
\caption{\label{fig:exampletod}Example timestreams from a single bolometer during a cloudy observation.
The top panel shows the Stokes $I$ (intensity) signal, and the middle panel shows demodulated Stokes $Q$ and $U$ polarization signals, with the offsets due to instrumental polarization. 
The bottom panel shows the slope of the simple linear regression between $I$ and $Q$, which corresponds to the signed $Q$ polarization fraction.}
\end{figure}

\cref{fig:exampletod}%
shows an example of the bolometer timestreams from a four-hour observation on December 18, 2014.
During the observation, the telescope is azimuthally scanning the sky back and forth in an azimuth range of $133^\circ$--$156^\circ$ at a constant elevation of $30^\circ$.
The precipitable water vapor (PWV) increases from $\sim0.8\,\mathrm{mm}$ to $\sim1.6\,\mathrm{mm}$ during the time of this data set.
The PWV is provided by the APEX experiment~\citep{Gusten2006AA} using a commercial LHATPRO microwave radiometer~\citep{Rose2005AtmRe}.
The APEX PWV will be partially correlated with the PWV at the \textsc{Polarbear} site $6\,\mathrm{km}$ away.

The intensity signal shown in the top panel is continuously fluctuating by $\pm3\,\mathrm{K}$ like a random walk, which is due to atmospheric turbulence.
The Stokes $Q$ in the middle panel, on the other hand, has negatively directed, burst-like structures, down by as much as $\sim0.3\,\mathrm{K}$ relative to an offset $A_0\sim0.2\,\mathrm{K}$.
The Stokes $U$ also shown in the middle panel has much smaller variation than $Q$, which means that the burst-like signal is horizontally polarized.
This property agrees with the expectation for the cloud signal that we described in \autoref{subsec:polarizationmodel}.

The bottom panel shows the slope of the simple linear regression between the intensity and $Q$ polarization signals, i.e.\ the signed $Q$ polarization fraction.
The intensity signal is a combination of the clouds and atmosphere, whose contributions are both a few Kelvin.
However, the time scale of the polarized-burst signal is shorter than that of the atmosphere.
Thus, in this calculation, we apply a simple high-pass filter by subtracting the baseline for each $\sim$50-second one-way scans and minimize the contribution of the atmosphere.
In the absence of the bursts in $Q$, the polarization is $\sim 0.1\%$, which is consistent with the level expected solely due to instrumental intensity-to-polarization leakage and no atmospheric polarization~\citep{Takakura2017JCAP}. 
On the other hand, the polarization fraction significantly increases to $5$--$10$\% at the timings of the bursts.
The polarization fraction of $10$\% at the elevation of $30^\circ$ is larger than the estimate for the spherical case as \cref{eq:polfrac} but can be explained by horizontally aligned column or plate crystals as shown in \cref{fig:polfrac_vs_shape}.

\begin{figure*}\centering
\begin{tabular}{c}
\includegraphics[width=0.5\textwidth]{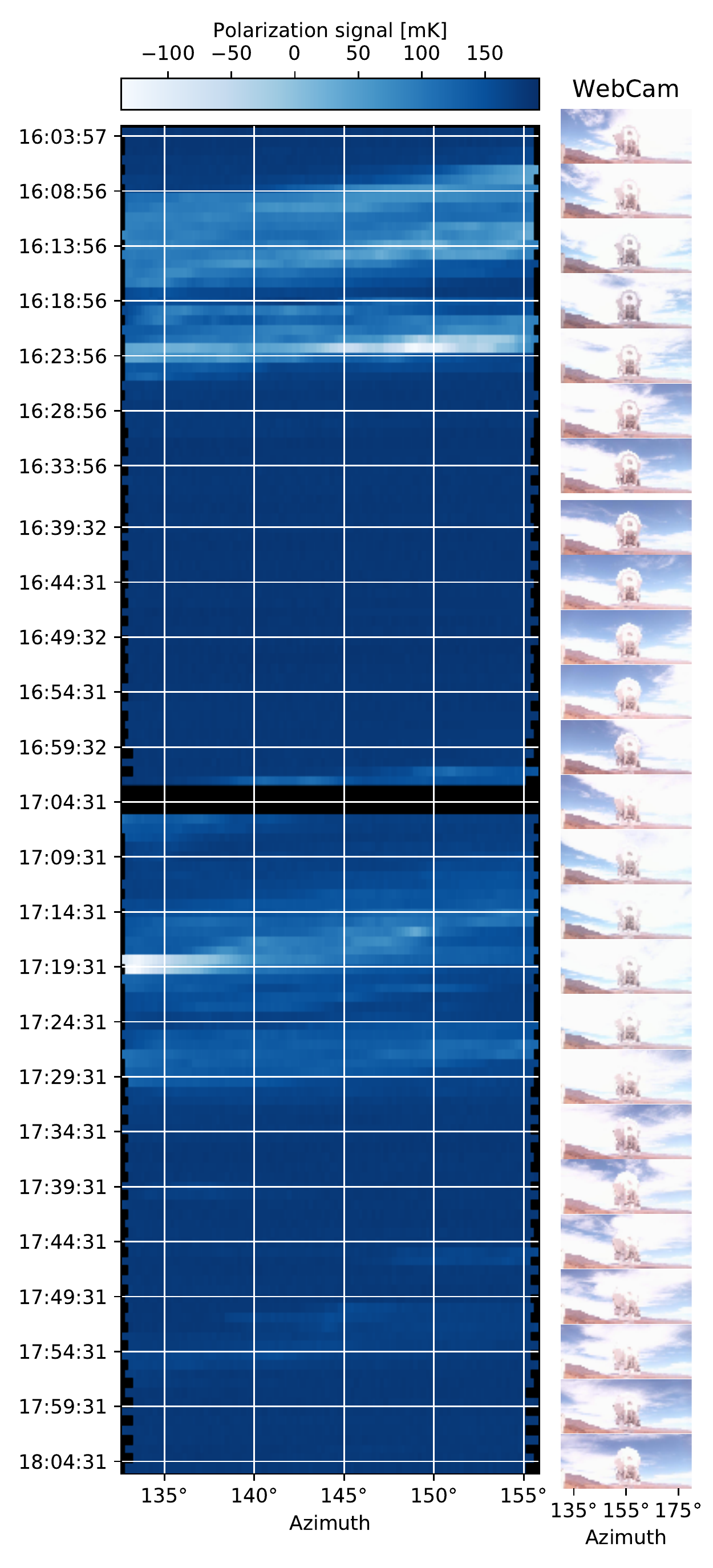}
\includegraphics[width=0.5\textwidth]{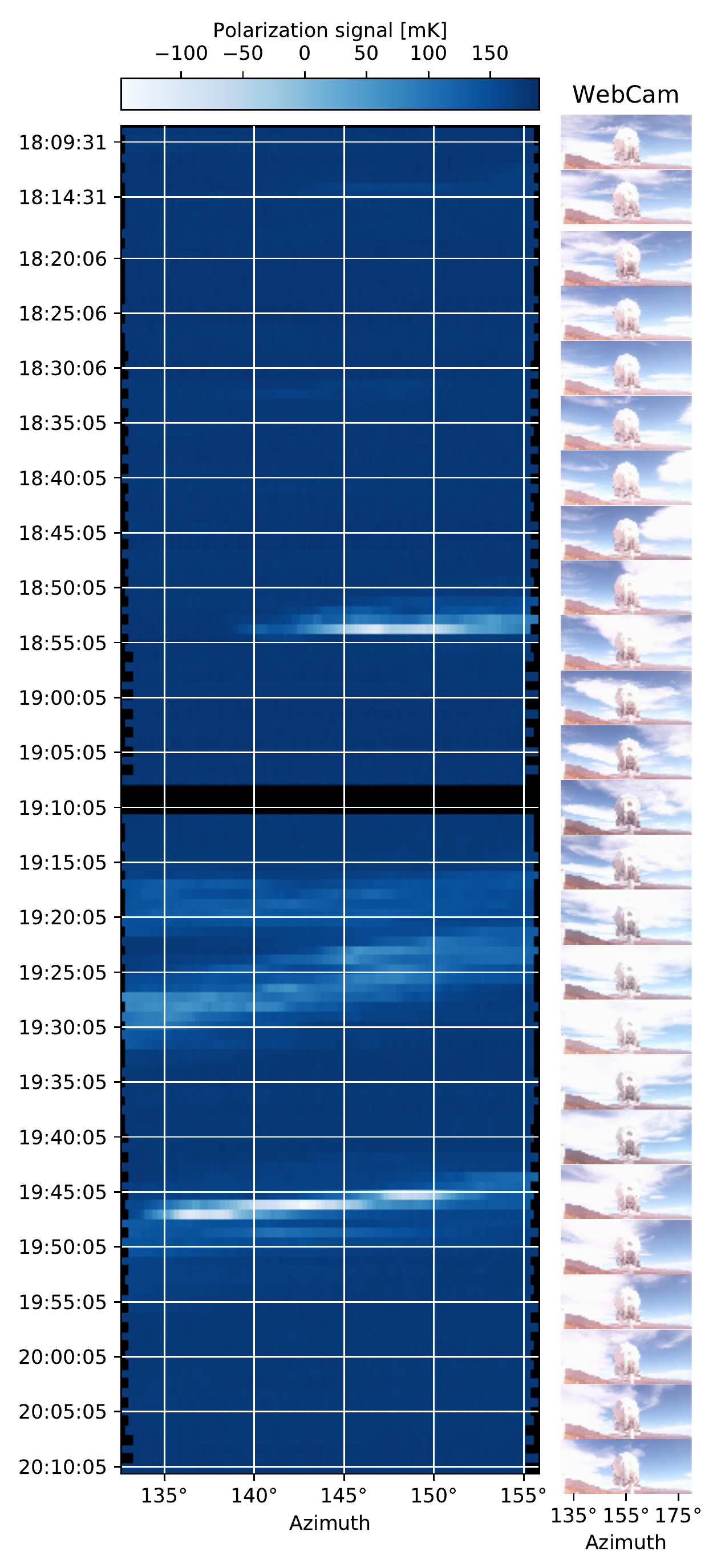}
\end{tabular}
\caption{\label{fig:examplebolocam}Comparison of the bolometer timestream with the webcam.
The Stokes $Q$ polarization timestream is mapped as a function of the azimuth ordered by time and shown as color.
There are gaps to tune the instruments every one hour, which are shown as black.
The photos of the webcam are shown next to the map at the corresponding times.
The azimuth range of the scan roughly corresponds to the left half in each of the photos.
The photos show that the white clouds are carried by the wind across the sky from the right to the left.
The polarized-burst signals in the map coincide with the photos with clouds and have the same trend from the right to the left.}
\end{figure*}

\cref{fig:examplebolocam} is another illustration of the same data, which is the map of the $Q$ polarization data as a function of the azimuth and time accompanied with snapshots from the webcam.
Each horizontal row of the map corresponds to each leftward or rightward scan that takes 50 seconds.
Around the time of 19:45:05, there is a structure within the row, which means that the variation of the polarized-burst signal is more rapid than the scan time.
In the next two scans, the structure appears consistently but in different azimuth, which suggests that the source of the signal is moving within the scan area.
The time scale of the motion is several minutes.
The existence of the polarized-burst signals and their motion from right to left agree well with those of the clouds in the webcam.
This result also supports the argument that the origin of the signal is a cloud.
Note that the maximum elevation of the webcam FOV is $20^\circ$ and does not exactly cover the scanning elevation of $30^\circ$.
However, the size of the clouds in the photos are sufficiently large to cover most of the sky, thus we suppose that the clouds would expand to the line of sight of the telescope.

We have considered other possibilities to create the polarized-burst signal, but none of them can explain the data.
Sudden responsivity variations may couple to the steady instrumental polarization $A_0$ and cause apparent variations in the $Q$ timestream.
However, it cannot explain the variation of the $Q(t)$ to negative values in \cref{fig:exampletod} because the responsivity of the TES detector does not change its sign.
Besides, the $2f$ signal, which is another stable optical signal from the CRHWP, does not exhibit such variations.
Temperature variations of the primary mirror could change the instrumental polarization, but the polarization fraction of the effect is expected to be less than $0.1\%$~\citep{Takakura2017JCAP}.
The far sidelobe of the telescope may have larger polarization leakage and see the ground and another part of the sky.
However, the spurious signal from the ground should stay in the same azimuth, while that from the sky should be gradual rather than burst-like fluctuations.
Condensation and evaporation of water vapor on the primary mirror may also cause spurious polarization but would not happen on rapid enough timescales.

\subsection{Cloud Detection Using the Webcam}\label{subsec:webcamstat}
We analyze all of the photos from the webcam and obtain the statistics of the clouds at the \textsc{Polarbear} site, in the Atacama Desert in Chile.
Note again that the FOV of the webcam covers a small fraction of the sky and that the telescope points to a sky region outside of the FOV in all the observations.
The webcam images are not useful during the night.
We also removed the pictures taken at dawn or dusk in the following analysis because the gradient of the brightness in the sky increases false detection rate.

\begin{figure}\centering
\includegraphics[width=1.0\linewidth]{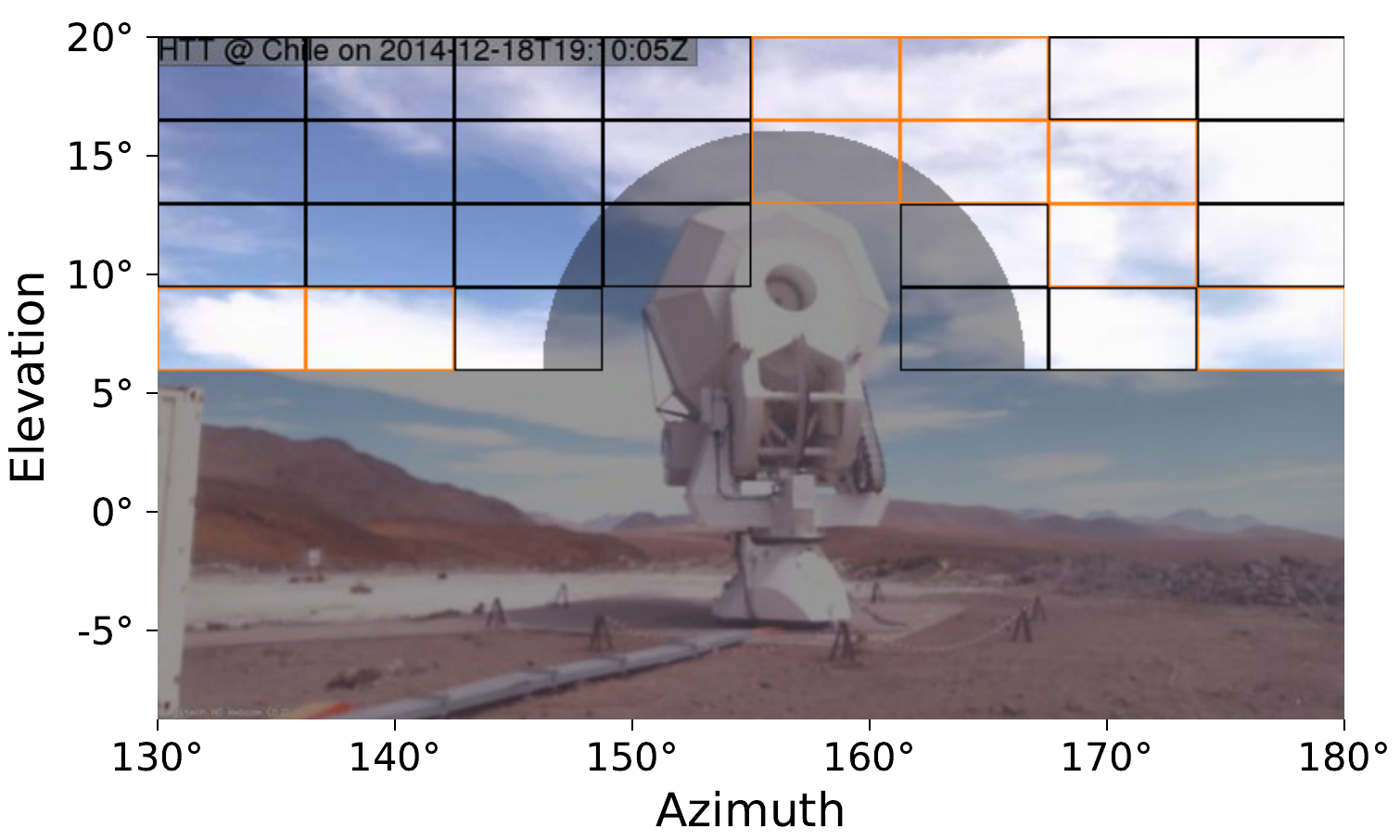}
\caption{\label{fig:webcampatch}Example of the cloud detection with a webcam image.
The shadowed region shows the mask, which is fixed for all the images.
The orange and black rectangles show patches with and without cloud detection in this image.}
\end{figure}

The basic idea of the cloud detection algorithm is to find white regions in the picture.
First, we mask the telescope, mountain, and ground, and split the sky into $29$ patches as shown in \cref{fig:webcampatch}.%
For each patch, we calculate the average Red-Green-Blue color $(R,G,B)$ among the pixels.
Then, we convert the color into Hue-Saturation-Value color $(H,S,V)$, specifically~\citep{HSV1978}
\begin{align}
S&=\frac{\max(R,G,B)-\min(R,G,B)}{\max(R,G,B)}\;,\\
V&=\max(R,G,B)\;.
\end{align}
We set the thresholds to detect the clouds in each patch as $S<0.1$ and $V<0.98$, where the former condition rejects the blue sky and the latter cuts saturated pixels due to the Sun.
The performance of the cloud detection is checked by eye for pictures from several days chosen randomly.
The algorithm often fails to detect faint clouds as in \cref{fig:webcampatch} but rarely make false detections, which are occasionally caused by ghost images.

\begin{figure}\centering
\includegraphics[width=1.0\linewidth]{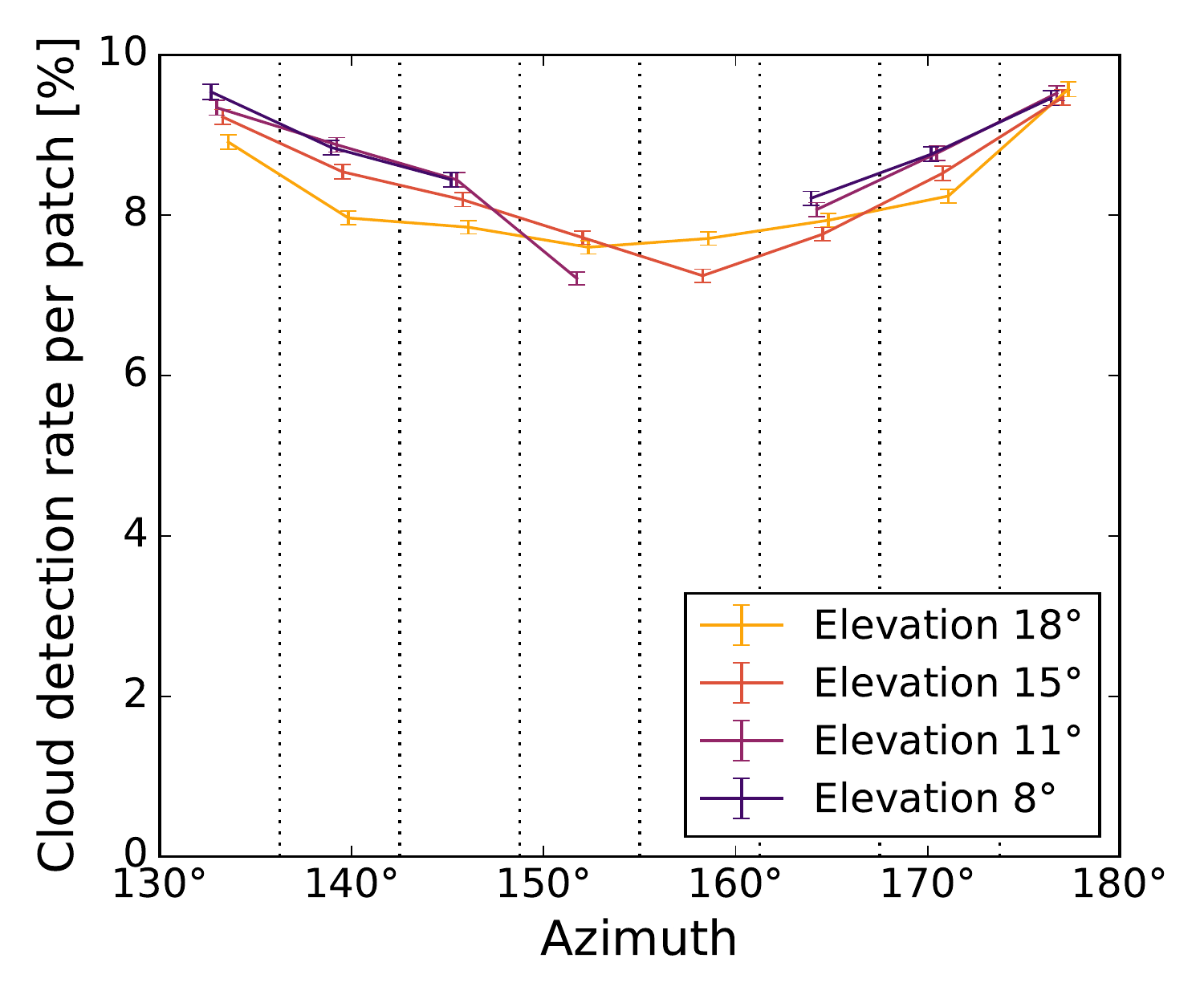}
\caption{\label{fig:patch_dependence}Cloud detection rate for each patch of the sky shown in \cref{fig:webcampatch}.}
\end{figure}

\cref{fig:patch_dependence}%
shows the rate of cloud detection for each patch, i.e.\ the number of shots with positive cloud detection in the patch divided by the total number of shots.
Although there is a small gradient in elevation, possibly due to the difference in the area of the sky, there is no clear tendency for clouds to appear in any particular region of the FOV. 
The clouds may be created outside of the webcam FOV, but they are expected to persist sufficiently long enough such that they will pass across the FOV.

\begin{figure}\centering
\includegraphics[width=1.0\linewidth]{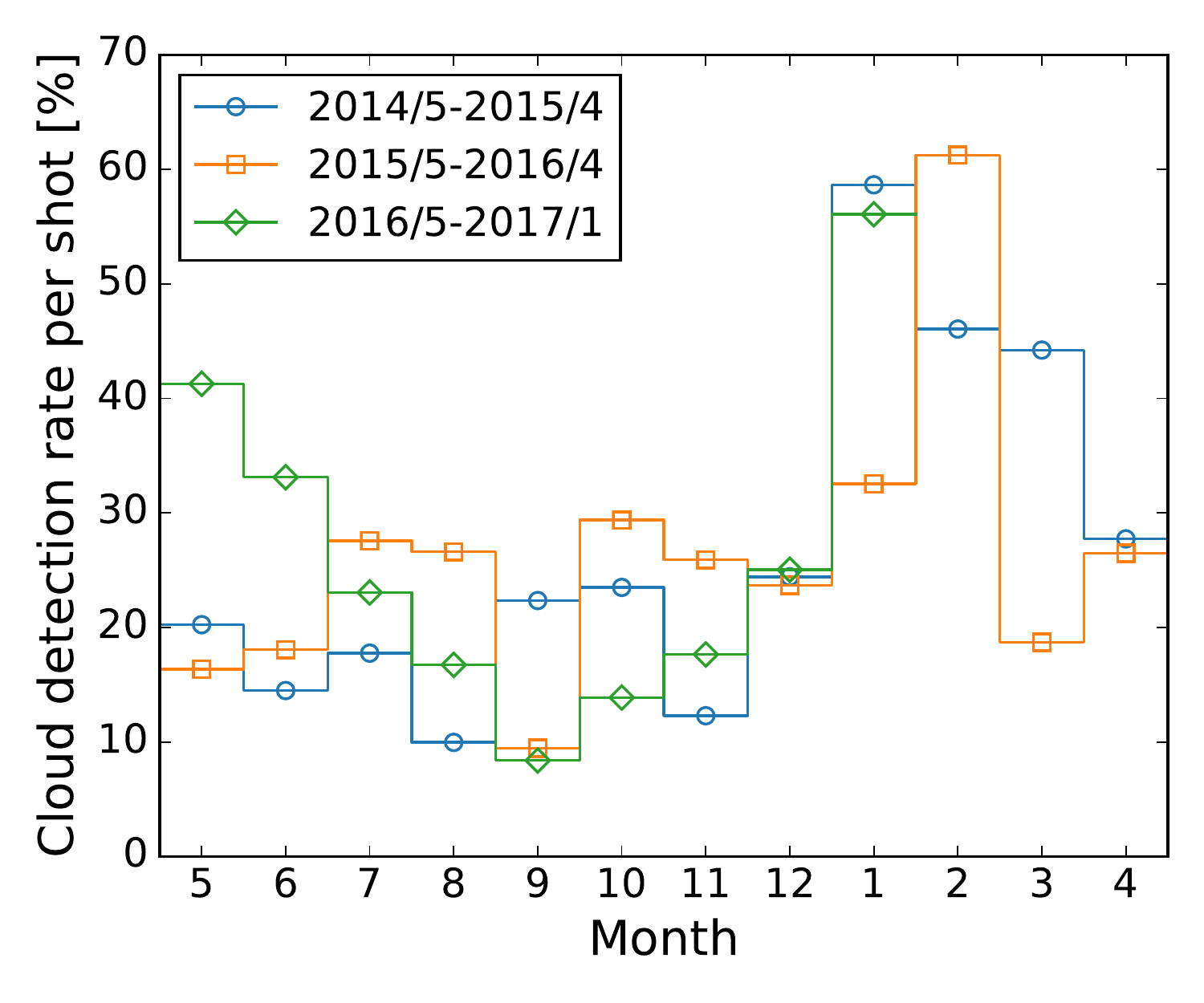}
\caption{\label{fig:annualvariation}Annual variation of the cloud detection rate.
Each line shows the result in the different year.
The cloud detection rate increases in January and February every year because of Altiplanic winter.}
\end{figure}

\begin{figure}\centering
\includegraphics[width=1.0\linewidth]{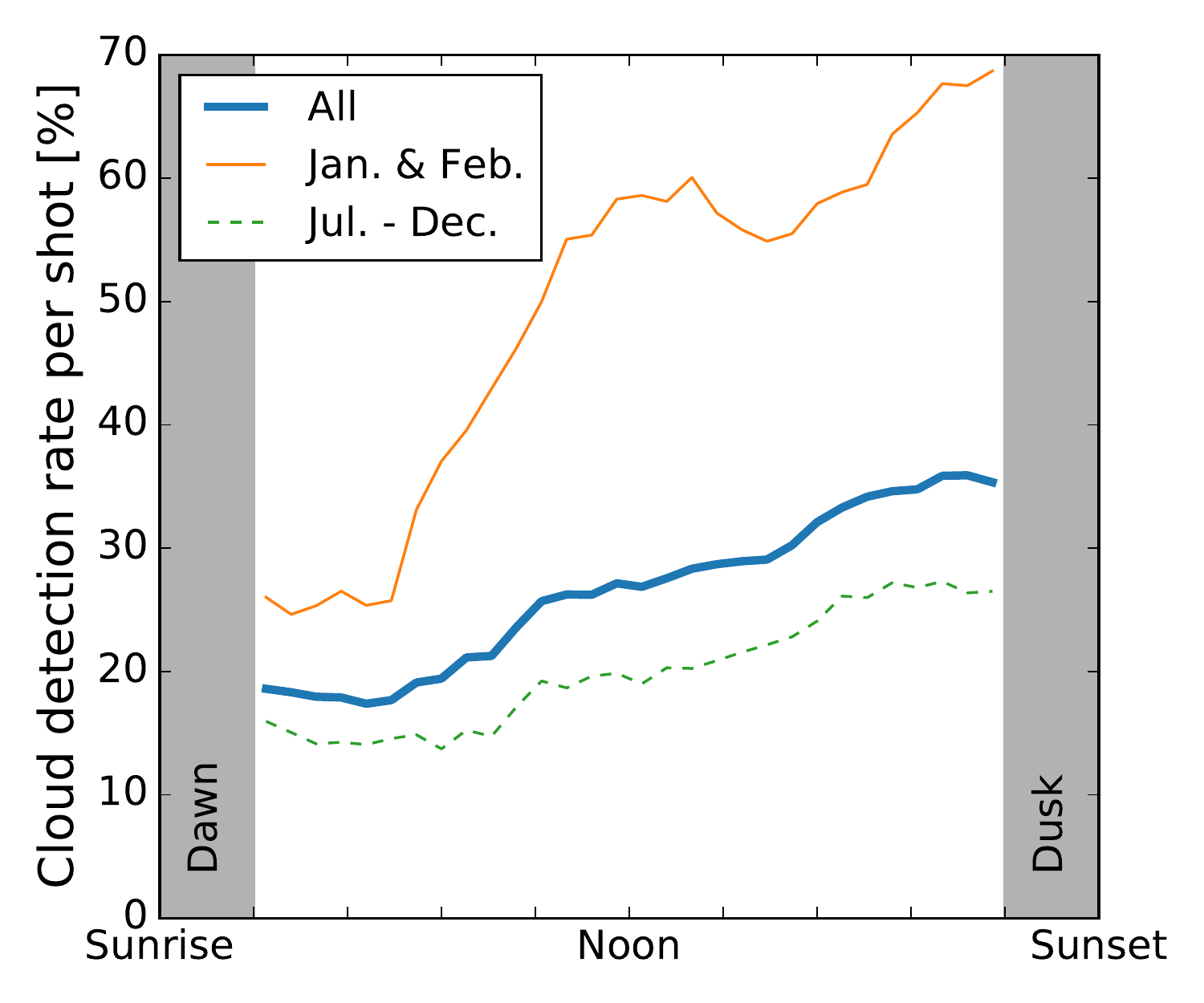}
\caption{\label{fig:diurnalvariation}Daytime variation of the cloud detection rate.
The horizontal axis represents the time in the daytime, which is scaled with respect to the sunrise and sunset times to adjust their seasonal variations.
The blue line shows the result from all the data, and the orange and green lines show the results in the high (January and February) and low (from July to December) seasons, respectively.
The shadowed regions show the mask used to cut dawn and dusk.}
\end{figure}

\cref{fig:annualvariation,fig:diurnalvariation}%
show annual and daytime variations of the cloud detection rate per shot.
Here we flag a shot as ``cloudy'' when a cloud is detected in at least one patch according to the algorithm specified above.
In \cref{fig:annualvariation}, we can see the significant cloud detection rate increase around February, which is known as Altiplanic winter.
During that season, the cloud detection rate seems to increase in the afternoon (\cref{fig:diurnalvariation}).
It might be due to the lift of the atmosphere heated up by thermal conduction from the ground, which is also heated by the sunlight~\citep{Erasmus2001CTIO}.
During July through December, on the other hand, the daytime trend is moderate.

\begin{figure}\centering
\includegraphics[width=1.0\linewidth]{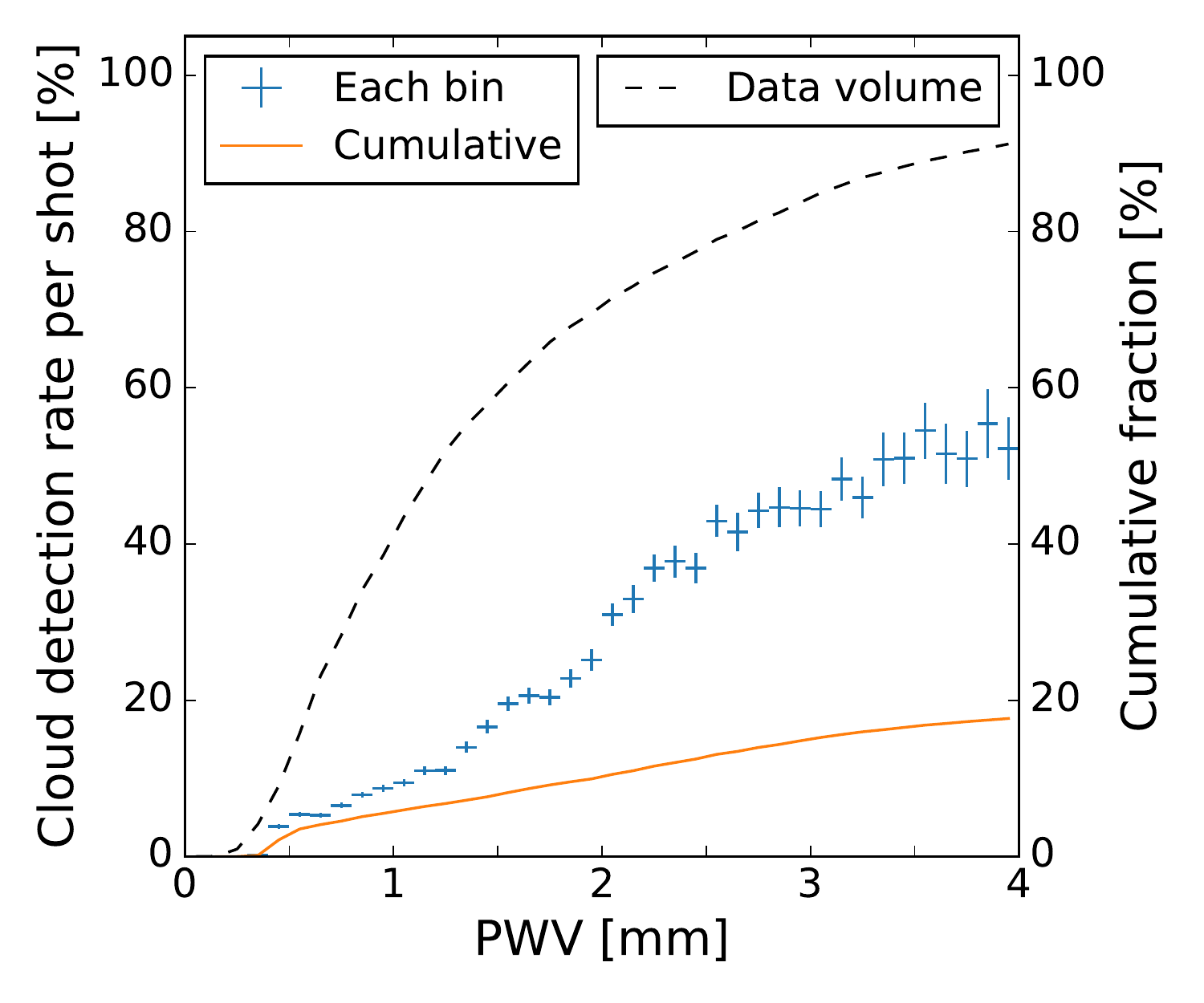}
\caption{\label{fig:pwvdependence}The dependence of the cloud detection rate on the PWV provided by the APEX experiment.
Each blue point shows the rate for data in each PWV bin and the orange line shows the rate for all the data below the PWV.
The cumulative fraction of the data volume is also shown as the black dashed line.}
\end{figure}

\cref{fig:pwvdependence}%
shows the correlation between the cloud detection rate per shot and the APEX PWV.
There is a clear correlation between the cloud detection and PWV.
This is an expected trend and provides additional supports and validation to our cloud detection method.

The overall cloud detection rate per shot is about $26\%$.
Note that we use only the daytime photos and have no information during the night.

\subsection{Cloud Detection in the Bolometer Data}\label{subsec:bolostat}
By using the polarized-burst signals in the bolometer data, we also detect clouds as shown in \cref{fig:exampletod}.
We use the \textsc{Polarbear} data from July 2014 to January 2017.
The cloud detection results are compared with the webcam study described in \autoref{subsec:webcamstat}.

Since the typical size of the clouds should be larger than the FOV of \textsc{Polarbear}, $3^\circ$, the cloud signal is correlated among all the detectors.
Thus, we can improve the sensitivity to detect the clouds by averaging the timestreams among all the detectors.
Similar to \cite{ABS2014Demod}, we separate the averaged polarization timestreams, $Q(t)$ and $U(t)$, into two components using the principal component analysis (PCA) method as
\begin{equation}\label{eq:PCA}
X_1(t) + i X_2(t) = \{Q(t)+iU(t)\}\,e^{-i\phi}\;,
\end{equation}
where the rotation $\phi$ is determined to maximize the variance of $X_1(t)$.
Since the secondary component $X_2(t)$ is dominated by the detector white noise,
the signal-to-noise ratio of the cloud is calculated as 
\begin{equation}\label{eq:SNR}
\mathrm{SNR} = \sqrt{\frac{\sigma(X_1)^2}{\sigma(X_2)^2}-1}\;,
\end{equation}
where the $\sigma$ denotes the standard deviation.
The polarization angle of the signal $\psi$ is obtained as
\begin{equation}\label{eq:polangle}
\psi=\frac{\phi}{2}\;.
\end{equation}
Here, we assume that the signal is almost horizontal as shown in \cref{fig:exampletod}, and constrain $\pi/2<\phi<3\pi/2$, which cannot be determined by PCA due to degeneracy.

\begin{figure}\centering
\includegraphics[width=1.0\linewidth]{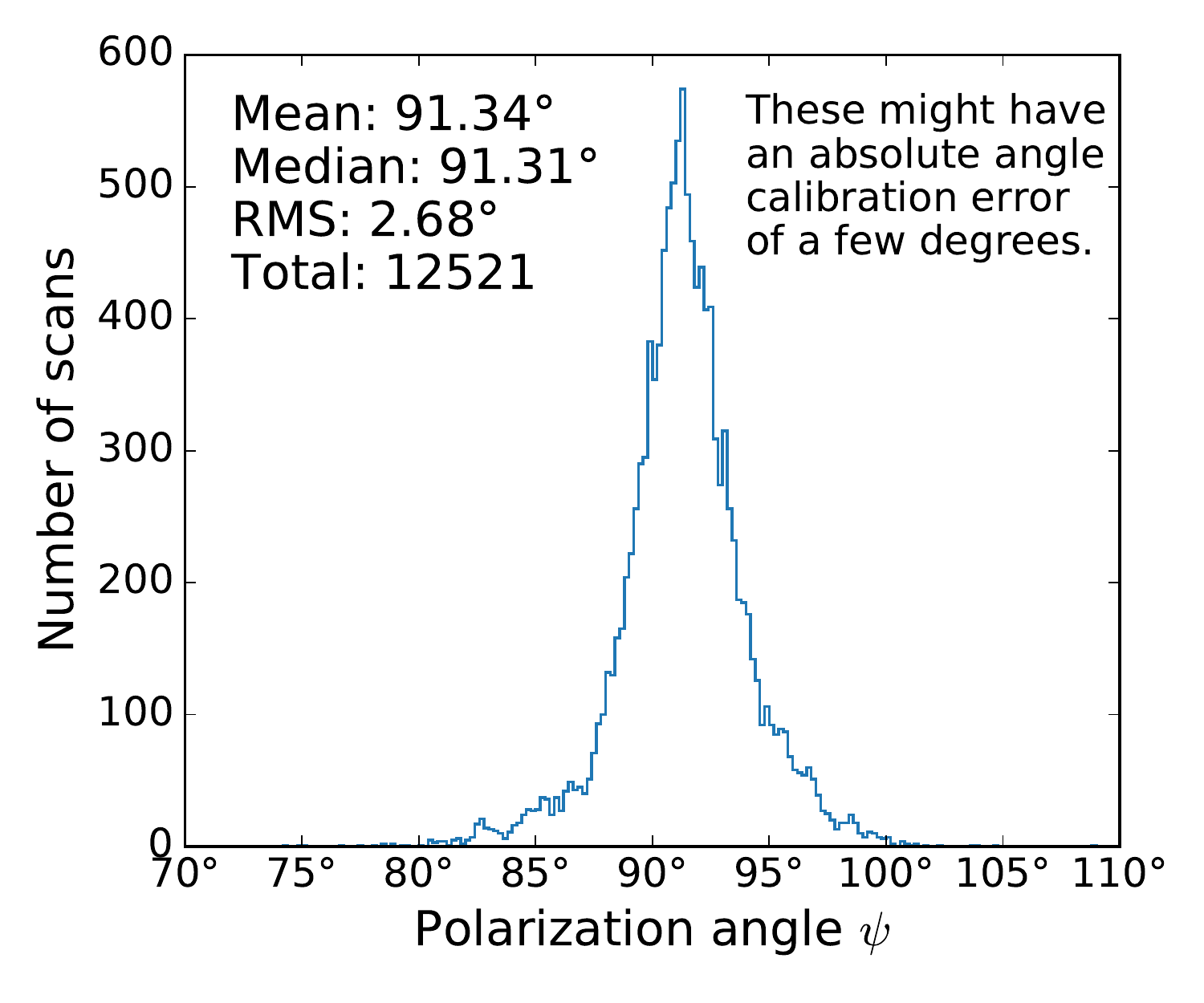}
\caption{\label{fig:polanglehist}The histogram of the polarization angle of the polarized-burst signal $\psi$.
The mean, median, and root-mean-square (RMS) values are calculated from samples in the range of $70^\circ<\psi<110^\circ$.
There are no other bunches outside of the range shown.
Note that we have not finalized the absolute angle calibration for this data set yet, and it might have an error of a few degrees.}
\end{figure}

\cref{fig:polanglehist}%
shows the histogram of the polarization angle of the polarized-burst signal $\psi$ with $\mathrm{SNR}>10$ for each scan.
If we have any instrumental noise sources other than the clouds, e.g.\ the intensity leakage due to the detector non-linearity, the polarized ground structure, and the HWP encoder error, they could appear at angles $\psi\nsim90^\circ$.
Having only the single peak at $\psi\sim90^\circ$ supports that such extra noises are not significant and that all the polarized-burst signals are most likely coming from clouds.
There is still a possibility of the responsivity variation due to electrical noise, but it cannot explain the coincidence with the webcam described below.
Both the width of the peak by $\mathrm{RMS}=2.68^\circ$ and the offset from $\psi=90^\circ$ might be due to a systematic error of the cloud signal or the instrument, which is still under investigation (see more discussions in \autoref{subsec:polcal}).

Each of the CMB observations typically takes one hour and contains $40$--$70$ left and right scans at a constant elevation.
We calculate the SNR and polarization angle $\psi$ of the cloud signal for each scan and take the values with the highest SNR as representative of the CMB observation.
On the other hand, the detection of clouds by the webcam is determined by a detection in at least one of the ${\sim}12$ photos taken during the observation.
Note that the CMB observations are performed at all times of the day, but the webcam can be used during the daytime only.

\cref{tab:cesfraction}%
\begin{table}
\centering
\begin{tabular}{|l|l||rr|}
\hline
\multicolumn{1}{|l|}{Data} & \multicolumn{1}{l||}{Webcam} & \multicolumn{2}{c|}{Polarized burst}\\\hline\hline
Daytime & All & 16.1\% & $(295/1835)$ \\
& Cloud & 46.3\% & $(279/602)$ \\
& No cloud & 1.3\% & $(16/1233)$ \\\hline
Night & \rule[0.7ex]{4.0em}{0.1pt} & 9.7\% & $(458/4735)$\\\hline
\end{tabular}
\caption{\label{tab:cesfraction}Fractions of the one-hour observations containing polarized-burst signals found in the bolometer data for the data sets corresponding either to a concurrent cloud detected in the webcam or to no cloud detected.
The fraction of burst-like signals occurring at night is also listed, though clouds cannot be identified by the webcam during the night.
The original numbers of one-hour observations are shown in parentheses.}
\end{table}%
shows the coincidence of the cloud detection in the bolometer data with that in the webcam.
For the data with the clouds in the webcam, the rate of the polarized-burst detection significantly increases to $46.3\%$ compared to $1.3\%$ for the data without the clouds.
Assuming that the appearance of polarized bursts in the bolometer data and the appearance of clouds in the webcam each occur at their observed rates but are also independent, then the significance of observing such a strong covariance between them is estimated to be $> 24\,\sigma$.
Note again that the FOV of the webcam does not cover the sky regions of the CMB observations.
That could be the reason for the deviation from the perfect separation.
Also, the daytime rate of the polarized burst is $16.1\%$, which is smaller than the ${\sim}30\%$, the fraction of data with clouds detected by the webcam (\autoref{subsec:webcamstat}).
It is because CMB observations are not performed when $\mathrm{PWV}>4\,\mathrm{mm}$, or the webcam is more sensitive to clouds than bolometers in this analysis using the thresholds above.
For the night data, we have no information on the clouds from the webcam, but the polarized-burst signals are detected in $9.7\%$ of the data. 

\begin{figure}\centering
\includegraphics[width=1.0\linewidth]{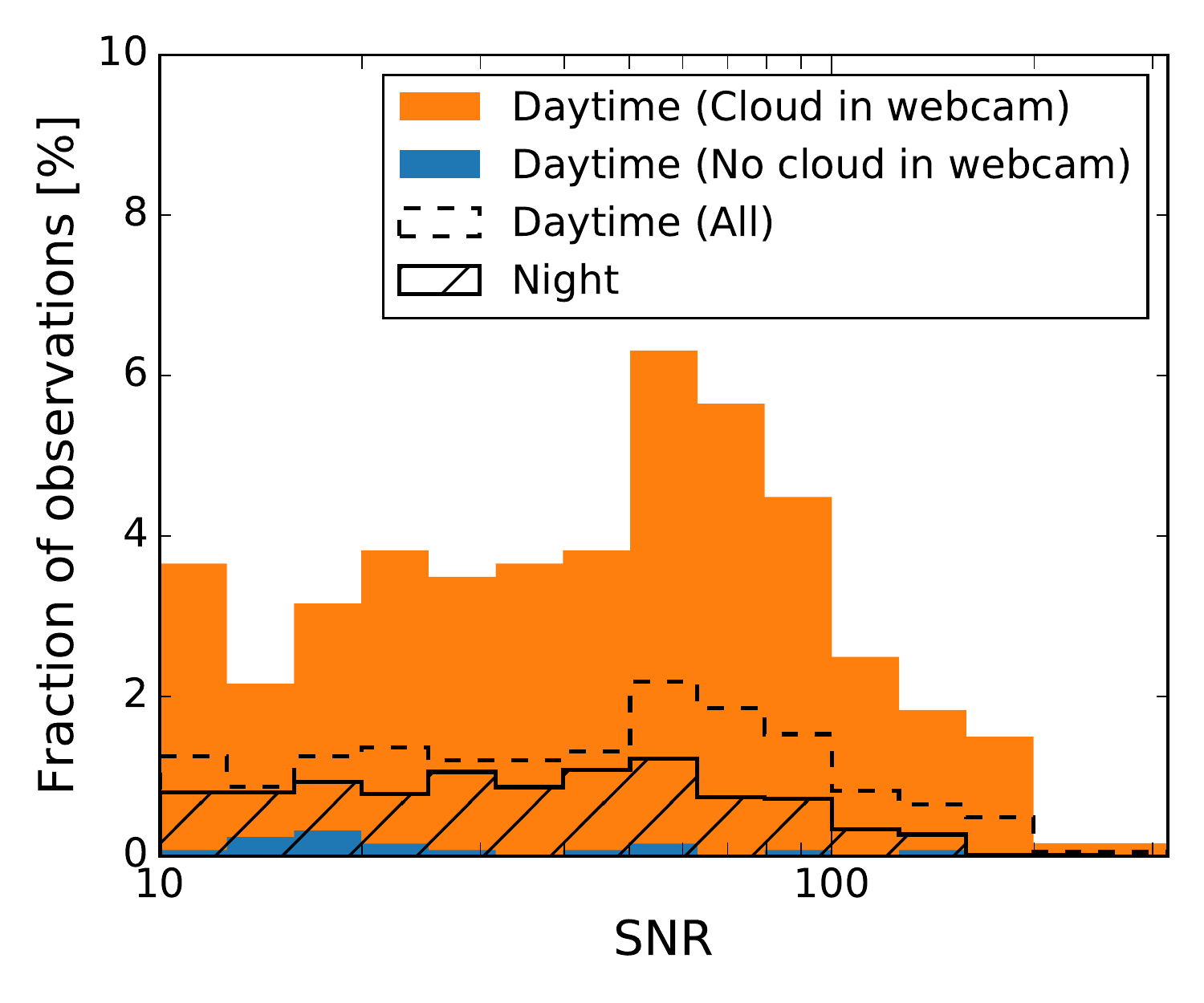}
\caption{\label{fig:noise_busrt_hist}The histogram of the maximum SNR of the polarized-burst signal in each observation normalized by the number of the observation.
The blue and orange histograms show the results for the daytime data with and without clouds detected by the webcam, and the result for all the daytime data is shown as the dashed line.
The result for the night data is also shown as the hatched histogram.
The outliers in the blue histogram might be due to clouds outside of the webcam FOV.}
\end{figure}

\cref{fig:noise_busrt_hist}%
shows the distribution of the SNR for each data set.
Note that it shows the partial data with high SNR (\cref{tab:cesfraction}).
Again, it shows the clear difference between the data with and without clouds detected by the webcam.
Besides, the SNR distribution for the night data is very similar to that for the daytime data with clouds.
As explained in \autoref{subsec:polarizationmodel}, the clouds scatter the thermal emission from the ground, which takes place also at night.
Our results support the expectation that the cloud signal also exists during the night.

\section{Discussion} \label{sec:discussion}
In the previous section, we have shown that the polarized cloud signals are detected in the CMB observations with \textsc{Polarbear}.
Here, we discuss the impact of the clouds on CMB polarization measurements.

\subsection{Degradation of the Statistical Precision}\label{subsec:efficiency}
CMB measurements require thousands of hours of observations with high-sensitivity detectors to measure the faint CMB signals.
The cloud signal is just noise that lowers the quality of the observations.

One simple approach to reduce the impact of the clouds is to drop noisy data, but that would inevitably reduce the observation efficiency.%
\footnote{Since the cloud signal only affects the Stokes $Q$ polarization, it may be possible to save the Stokes $U$ component.}
It is possible to detect cloud signals in the bolometer data to an extent as performed in \autoref{subsec:bolostat}.
We find cloud-like polarized-burst signals in 16.1\% (9.7\%) of one-hour observations (\cref{tab:cesfraction}) and in 5.2\% (3.0\%) of leftward or rightward scans during the daytime (night).
One could use image analysis similar to \autoref{subsec:webcamstat}, in which we detect clouds in 26\% of all the daytime shots of the webcam.
As shown in \cref{fig:pwvdependence}, the fraction of data without clouds improves if we observe only in good PWV.
But, this decreases the total amount of data.

Even with data cuts, we expect residuals from faint clouds below the detection threshold.
These residuals will be present at low frequency with a $1/f$ behavior and degrade the detector performance at large angular scales.
Such residuals cannot be mitigated by polarization modulation techniques such as the CRHWP used in \textsc{Polarbear}.

If we detect the clouds using the bolometer data as performed in \autoref{subsec:bolostat}, the residual noise level will depend on the SNR threshold of the cloud detection.
For example, to achieve an SNR of $10$, the power of the cloud signal should be $100$ times larger than that of the detector noise.
In other words, the residual $1/f$ noise below the threshold could na\"ively degrade the sensitivity of the CMB angular power spectra by $100$ times in the worst case.
Tightening the threshold will mitigate the contamination but also decrease the observation efficiency.
Thus, optimization of the threshold is required to maximize performance.

The effect of clouds would depend on the telescope FOV and the detector beam size.
Small-aperture telescopes with a large FOV have a high probability of seeing clouds.
Large-aperture telescopes with small beams have a high instantaneous SNR of a cloud, because many detectors simultaneously observe the same cloud, which is larger than the beam size.

These studies could inform forecasting and optimization of future CMB experiments, such as CMB-S4~\citep{CMBS4}.

\subsection{Systematic Errors on CMB Measurements}\label{subsec:systematics}
Systematic errors due to the residual cloud signal are also of concern for CMB measurements.

The cloud signal is horizontally polarized, i.e.\ $Q<0$, and not symmetric fluctuations between plus and minus.
This highly non-Gaussian fluctuation could lead to possible systematics in the map as many map-making algorithms assume Gaussianity for noise fluctuations.
This systematics can be mitigated to some extent by parallactic angle rotation.

The cloud signal would affect foreground removal due to its distinct frequency dependence.
By using maps at multiple frequencies, we separate the CMB and the other foregrounds that are stationary in the sky, i.e.\ not associated to any Earth or atmospheric motion, such as the Galactic dust and synchrotron emissions.
Since the cloud signal has the Rayleigh scattering spectrum as $\propto \omega^6$ (see \cref{eq:opticaldepth} with additional $\omega^2$ from the spectrum of the ground emission), it would appear rising in frequency, similar to a dust component (approximately $\omega^{1.5}$), but much steeper.
On the other hand, the atmospheric motion would likely de-correlate with astrophysical foregrounds, making the control of the clouds appropriate for a time domain analysis, rather than in maps.

Clouds staying in the same position may also cause systematic errors.
However, there is no significant localized cloud feature appearing in \cref{fig:patch_dependence}.
In addition, any localized clouds would be fixed to ground structures, such as mountains,
but the rotation of the sky would change the relative positions and suppress systematic errors.

The daytime trend of the cloud detection rate shown in \cref{fig:diurnalvariation} might cause a systematic difference in the additional noise from the residual cloud signals between the morning and afternoon observations.
However, the yearly motion of the Earth gradually shifts the observing time of the CMB patch fixed on the sky.
For yearlong observations, the difference will be at least partially averaged.

Performing null tests sensitive to the clouds is necessary to validate the data.
A rising vs.\ setting split can test for localized clouds, and a summer vs.\ winter split can test the impact of the diurnal variation as mentioned above.
A low-PWV vs.\ high-PWV split can test the cloud rate because of their correlation as shown in \cref{fig:pwvdependence}.

\subsection{Polarization Angle Calibration}\label{subsec:polcal}
While ice clouds are a nuisance in CMB observations, the polarized signal from the clouds could be a useful calibrator for the absolute polarization angle.
As explained in \autoref{subsec:polarizationmodel} and demonstrated in \cref{fig:exampletod,fig:polanglehist}, the signal is horizontally polarized mainly because the column or plate ice crystals are aligned horizontally by gravity.
Cirrus clouds lie at an altitude of $\sim$$10\,\mathrm{km}$, so the distance from the telescope is sufficient to achieve a far-field measurement for \textsc{Polarbear} with the $2.5\,\mathrm{m}$ diameter aperture observing at $150\,\mathrm{GHz}$.
Furthermore, the clouds are diffuse objects, making beam systematics of less concern.
These two properties are better than near-field calibrators, such as the sparse wire grid in front of the telescope~\citep{Tajima2012JLTP,ABS2018}, which need a connection from the near-field measurement to the far-field beam relying on the optics model.
Although the spectrum of the cloud signal depends strongly on the observing frequency as $\propto \omega^6$, the polarization angle does not depend on the frequency.
This addresses an uncertain polarization angle rotation feature of Tau~A, a popular polarized celestial source.
Compared to calibrators on the ground, cloud calibration would make it possible to operate the detectors with typical sky loading, as opposed to extra loading observing a source near the ground, and it would not have uncertainty in extrapolating the pointing model of the telescope.

In \cref{fig:polanglehist}, the precision of the polarization angle calibration for each scan is only $2.7^\circ$, but the uncertainty of the mean value can be shrunk by accumulating statistics to $0.03^\circ$, provided the errors are independent and Gaussian distributed.
That is better than the statistical uncertainty of $0.16^\circ$ from the polarization angle calibration from nulling the apparent correlation between the CMB $E$ and $B$ mode patterns from two years of \textsc{Polarbear} data~\citep{POLARBEAR2017ApJ}.
In addition, the cloud polarization is absolutely referenced to gravity, and it does not use assumptions about the symmetry properties of the CMB.

On the other hand, the cloud signal may have its own systematic errors.
For example, wind and electrification may slightly tilt the ice crystals.
The ground emission may not be uniform due to local features, e.g.\ deserts, mountains, lakes, snowfields, etc.
The contribution of the Sun can become non-negligible.
We have estimated the systematic error by splitting the data into subsets for various observation conditions: year split, day-night split, scan-direction split, PWV split, outside-temperature split, and wind-speed split.
The median value for each subset has about $0.4^\circ$ variation.
However, that value also includes the systematic error of the instrument such as the time constant variation during the observation and imperfection of the pointing model.
Further investigation is necessary to separate them, but the possibility of having many measurements with various conditions demonstrates the potential usefulness of the cloud signal as a polarization angle calibrator.

\subsection{Prescription for Future Experiments}\label{subsec:prescription}
For future ground-based CMB experiments aimed at more precise measurement of CMB polarization, such as the CMB-S4, steps to mitigate the contamination of the cloud signal will be necessary.

One approach is \textit{in situ} measurement of the clouds.
In \autoref{subsec:example}, we have demonstrated a simple cloud detection technique using the webcam for monitoring the telescope.
Even with the limitation of its FOV, the significant coincidence of the cloud detections between the webcam and bolometers is observed as shown in \cref{tab:cesfraction,fig:noise_busrt_hist}.
It can be improved by using a whole-sky camera and a co-mounted infrared camera, which would be useful during the night~\citep[e.g.][]{Suganuma2007PASP}.
As already mentioned in \cite{Pietranera2007MNRAS}, the most informative but challenging method is polarized lidar~\citep[e.g.][]{Lewis2016JAtOT}, which shoots a laser pulse to the sky, receives the scattered light, and characterizes the atmospheric properties along the line of sight including the shape, size distribution, and orientation of the ice crystals.
These tools would enable reliable cloud detection and precise data selection.
This would also help with understanding the clouds and reducing the systematic error of polarization angle.

Another approach might be to perform foreground separation in the time domain.
The clouds are obviously the frontmost component of the foregrounds for CMB observations.
The cloud signal has frequency dependence markedly different from that of the CMB and the other astrophysical foregrounds, i.e., approximately $\omega^{6}$ as opposed to $\omega^{-3}$ for the synchrotron and $\omega^{1.5}$ for the dust.
Therefore, it would be possible to separate the cloud signal in measurements with multi-frequency bands.
Here, it is important to observe the same position at the same time among detectors with different frequency bands.
The multi-chroic detector technique used in e.g.\ the Simons Array experiment~\citep{SimonsArray2016SPIE} would be useful for that purpose.

Of course, satellite missions are the best solution to avoid the clouds.
Balloon-borne experiments may see clouds in the stratosphere.
But their impact would be small because the particle size of the stratospheric clouds is smaller than the tropospheric clouds.

\section{Summary}\label{sec:summary}
The ice crystals in tropospheric clouds scatter the thermal emission from the ground and produce horizontally-polarized signals.
Especially, the column and plate crystals should have the tendency to align horizontally with respect to the ground, which enhances the polarization fraction up to tens of percents.

In this study, we have presented the measurements of the clouds with the \textsc{Polarbear} experiment.
The horizontal polarization and the significant coincidence between the detectors and webcam strongly support the argument that the polarized-burst signals are actually coming from the clouds.
Note that the polarization modulation technique using the CRHWP is essential to perform the clear separation between the intensity and polarization signals.

Dropping data with clouds could decrease the efficiency of CMB observations.
In the webcam analysis, clouds are detected in $26\%$ of all the daytime shots.
In addition, the residual cloud signal may become a critical source of low-frequency noise and systematic error that cannot be mitigated with polarization modulation techniques.
In future experiments, \textit{in situ} measurements of the clouds with extra instruments or a sophisticated analysis combining multi-frequency detectors will help mitigate the contamination.

On the other hand, the cloud signal could potentially be a good calibrator of the absolute polarization angle with $0.03^\circ$ precision if the systematic errors of $0.4^\circ$ associated with it can be understood.

\acknowledgments
S.~Takakura was supported by Grant-in-Aid for JSPS Research Fellow JP14J01662 and JP18J02133.
The \textsc{Polarbear} project is funded by the National Science Foundation under Grants No.~AST-0618398 and No.~AST-1212230.
The James Ax Observatory operates in the Parque Astron\'omico Atacama in Northern Chile under the auspices of the Comisi\'on Nacional de Investigaci\'on Cient\'ifica y Tecnol\'ogica de Chile (CONICYT). 
The APEX PWV was obtained from the APEX Weather Query Form of the ESO Science Archive Facility.
This research used resources of the Central Computing System, owned and operated by the Computing Research Center at KEK, the HPCI system (Project ID:hp150132), 
and the National Energy Research Scientific Computing Center, a DOE Office of Science User Facility supported by the Office of Science of the U.S.\ Department of Energy under Contract No.~DE-AC02-05CH11231.
Kavli IPMU was supported by World Premier International Research Center Initiative (WPI), MEXT, Japan.
This work was supported by MEXT KAKENHI grant Nos.~21111002, JP15H05891, and JP18H05539, JSPS KAKENHI grant Nos.~JP26220709, JP26800125, and JP16K21744, and the JSPS Core-to-Core Program.
MA acknowledges support from CONICYT UC Berkeley-Chile Seed Grant (CLAS fund) Number 77047, Fondecyt project 1130777 and 1171811, DFI postgraduate scholarship program and DFI Postgraduate Competitive Fund for Support in the Attendance to Scientific Events.
CB, NK, and DP acknowledge support from the RADIOFOREGROUNDS project (radioforegounds.eu), funded by the European Commission's H2020 Research Infrastructures under the Grant Agreement 687312, the INDARK INFN Initiative and the COSMOS network of the Italian Space Agency (cosmosnet.it).
FB and CR acknowledges support from an Australian Research Council Future Fellowship (FT150100074).
D.~Boettger gratefully acknowledges support from grant ALMA CONICYT 31140004.
GF acknowledges the support of the CNES postdoctoral program.
Work at LBNL was supported by the Laboratory Directed Research and Development Program of Lawrence Berkeley National Laboratory under U.S.\ Department of Energy Contract No.~DE-AC02-05CH11231.
AK acknowledges the support by JSPS Leading Initiative for Excellent Young Researchers (LEADER).
FM acknowledges JSPS International Research Fellowship (Grant number JP17F17025).

\bibliographystyle{aasjournal}
\bibliography{reference}
\end{document}